# Project Lyra: Catching 1I/'Oumuamua – Using Laser Sailcraft in 2030

Adam Hibberd, Andreas M. Hein

Initiative for Interstellar Studies (i4is)

27/29 South Lambeth Road, London, SW8 1SZ, United Kingdom

## Abstract

Discovered in October 2017, the interstellar object designated 1I/'Oumuamua was the first such object to be observed travelling through our solar system. 1I/'Oumuamua has other characteristics never seen before in a celestial body and in-situ observations and measurements would be of extraordinary scientific value. Previous studies have demonstrated the viability of spacecraft trajectories to 'Oumuamua using chemical propulsion with a Solar Oberth burn at a perihelion as low as a few solar radii. In addition to chemical propulsion, there is also the possibility of missions involving light sails accelerated by the radiation pressure of a laser beam from a laser located on Earth. Based on a scaled-down Breakthrough Starshot beaming infrastructure, interplanetary missions and missions to the outer solar system have been proposed using lower sailcraft speeds of 0.001c relaxing the laser power requirements (3-30 GW for 1-100 kg spacecraft) and various other mission constraints. This paper uses the OITS trajectory simulation tool, which assumes an impulsive $\Delta V$ increment, to analyze the trajectories which might be followed by a sailcraft to 'Oumuamua, with a launch in the year 2030 and assuming it has already been accelerated to a maximum speed of 300km/s (approx 0.001c) by the laser. A minimum flight duration of 440 days for a launch in July 2030 is found. The intercept would take place beyond 82 AU. We conclude that the possibility of launching a large number of spacecraft and reaching 1I much faster than chemical propulsion would circumvent several disadvantages of previously proposed mission architectures.

## 1. Introduction

The first interstellar object to be observed within our Solar System is 1I/'Oumuamua, which was discovered in October 2017[1–3]. Research on this object is still ongoing and covers topics such as its shape [4–9], its composition [4,5,10–17], itsorigin[18–26], explanation for an observed acceleration [27–30], and estimates for the population of similar objects [2,5]. In 2018, a second interstellar object, 2I/Borisov was discovered [11,31,32]. However, while 2I/Borisov has characteristics which resemble solar system comets[33–35], it seems that 1I/'Oumuamua is unique in several respects such as its shape and composition. Hence despite the discovery of a second interstellar object, 1I remains a unique target for further exploration. The feasibility of missions to interstellar objects in general have been assessed in [36] and specific missions to 1I in [37,38]. [37,38] conclude that missions to 1I are feasible, using existing technologies, or technologies currently under development and could be launched several decades into the future, leaving enough time for development. In essence, this would include one or more planetary flybys and a solar Oberth maneuver. However, one of the caveats of these proposals is that they would take decades to reach the target and it is unclear whether the probe would be able to detect the object with sufficient precision. One way to omit such long trip times is to use alternative propulsion systems for reaching much higher velocities. Hein et al. [37] briefly covered the possibility of launching spacecraft propelled by laser sail propulsion, using a smaller version of the Breakthrough Starshot beaming infrastructure [39]. The spacecraft would be gram- to kilogram sized. A more recent paper by Turishev et al. [40] proposed the use of such an architecture for exploring a variety of outer solar system bodies, such as the conjectured Planet 9 [41–43]. However, these publications do not go into details of mission design.

This paper analyses trajectories a sailcraft might take to intercept 1I/'Oumuamua. A high power laser beam located on Earth is directed onto the sail and the consequent change in velocity is assumed to be instantaneous.

## 2. Materials and Methods
.

The trajectories are calculated usingan adapted version of the Optimum Interplanetary Trajectory Software (OITS) developed by Adam Hibberd, based on a patched conic approximation. With the patched conic approximation, only the gravitational attraction of a celestial body is taken into consideration within its sphere of influence and the gravitational attraction of other bodies is neglected. The primary modification to the software is to allow optimization with respect to flight duration rather than $\Delta V$.

The line of research elaborated here involves direct trajectories from Earth to 'Oumuamua in the year 2030. Thus in the presence of the sun's gravitational influence, given a launch time at Earth and arrival time at 'Oumuamua, there are two possible direct trajectories, 'short way' and 'long way'. The change in true anomaly, $\Delta\theta$, is limited to $0 \leq \Delta\theta \leq \pi$ for short way which leads to $(2\pi - \Delta\theta)$ for long way. For the purpose of this investigation, the two trajectories are separated into prograde and retrograde with respect to Earth's orbital velocity. Thus if $V_S$ is the heliocentric velocity vector of the s/c after leaving Earth's sphere of influence and $V_E$ is Earth's heliocentric velocity then $V_S.V_E > 0$ for prograde and $V_S.V_E < 0$ for retrograde. These trajectories are determined by solving the Lambert problem using the Universal Variable Formulation. The resulting non-linear global optimization problem with inequality constraints is solved applying the NOMAD solver[44]. The software repository holding OITS resides on github where a definition manual detailing the theory behind OITS can also be found [45].

The optimization is to minimize flight duration (T) assuming a constraint on maximum hyperbolic excess speed with respect to Earth of 300km/s (approx. 0.001c). This would translate into a beaming power of between 3 to 29 GW for spacecraft of 1 kg to 100 kg, respectively [40]. Furthermore a minimum limit on perihelion (Peri) is specified throughout as 3 Solar Radii from the centre of the sun, i.e. approximately 0.014AU.

## 3. Results

Refer to Figure 1 for the prograde solutions and Figure 2 for the retrograde solutions. In both Figures, two plots are shown: a 3D plot on the left and a 2D plot (looking down from above the ecliptic) on the right. The blue dashed line is Earth's orbit and the black lines are possible trajectories which a laser sail s/c could take. The intercept point with 'Oumuamua is at a great distance (>82 AU) so is excluded from the x/y/z axis scales chosen in these plots. As one might expect the prograde route is faster with mission durations of as low as T=440days for a launch in July 2030 (the precise optimum is July 1st).

Declination plots are also provided (Figure 3 for prograde and Figure 4 retrograde). The declination plots give an idea of the optimal latitude of the laser on Earth assuming one wishes to apply the beam radially outwards from the Earth's centre, which will also be approximately the minimum thickness of Earth's atmosphere. It can be observed that the optimal geocentric latitude of the beamer in July 2030 would be around 27° north. The constraint on Earth longitude for the beamer is less severe because the required RA/DEC values for the $\Delta V$ application do not change rapidly over the course of a day (they change on a time scale on the order of months). Candidate sites include locations in Florida, South Texas, the Sahara desert, North India and China. To make optimal use of the Earth's orbital velocity, the beam application would be at around 6am. Refer to Table 1 which gives beaming data for the 3 beaming scenarios mentioned in the Appendix of [40]. However if optimal beaming is contingent on operation at night then a date in October and an application at around midnight would be possible, but with a longer flight duration of 489 days. The latitude would be somewhere around 5° south of the equator.

**Table 1 Beaming Data**

| Number of minutes of beam application | Angle off zenith (degrees) assuming some agility in beaming direction & equatorial plane rotation | Approx. Beginning of beam application Apparent Solar Time[1] | Approx. Ending of beam application Apparent Solar Time[1] | Number of firings for Beaming duration =0.57days | Number of firings for Beaming duration =1.5days | Number of firings for Beaming duration =4days |
|---|---|---|---|---|---|---|
| 10 | 1.25 | 05:55:00 | 06:05:00 | 82.08 | 216.00 | 576.00 |
| 20 | 2.5 | 05:50:00 | 06:10:00 | 41.04 | 108.00 | 288.00 |
| 30 | 3.75 | 05:45:00 | 06:15:00 | 27.36 | 72.00 | 192.00 |
| 40 | 5 | 05:40:00 | 06:20:00 | 20.52 | 54.00 | 144.00 |
| 50 | 6.25 | 05:35:00 | 06:25:00 | 16.42 | 43.20 | 115.20 |
| 60 | 7.5 | 05:30:00 | 06:30:00 | 13.68 | 36.00 | 96.00 |
| 70 | 8.75 | 05:25:00 | 06:35:00 | 11.73 | 30.86 | 82.29 |
| 80 | 10 | 05:20:00 | 06:40:00 | 10.26 | 27.00 | 72.00 |
| 90 | 11.25 | 05:15:00 | 06:45:00 | 9.12 | 24.00 | 64.00 |
| 100 | 12.5 | 05:10:00 | 06:50:00 | 8.21 | 21.60 | 57.60 |
| 110 | 13.75 | 05:05:00 | 06:55:00 | 7.46 | 19.64 | 52.36 |
| 120 | 15 | 05:00:00 | 07:00:00 | 6.84 | 18.00 | 48.00 |
| 130 | 16.25 | 04:55:00 | 07:05:00 | 6.31 | 16.62 | 44.31 |
| 140 | 17.5 | 04:50:00 | 07:10:00 | 5.86 | 15.43 | 41.14 |
| 150 | 18.75 | 04:45:00 | 07:15:00 | 5.47 | 14.40 | 38.40 |
| 160 | 20 | 04:40:00 | 07:20:00 | 5.13 | 13.50 | 36.00 |
| 170 | 21.25 | 04:35:00 | 07:25:00 | 4.83 | 12.71 | 33.88 |
| 180 | 22.5 | 04:30:00 | 07:30:00 | 4.56 | 12.00 | 32.00 |
| 190 | 23.75 | 04:25:00 | 07:35:00 | 4.32 | 11.37 | 30.32 |
| 200 | 25 | 04:20:00 | 07:40:00 | 4.10 | 10.80 | 28.80 |
| 210 | 26.25 | 04:15:00 | 07:45:00 | 3.91 | 10.29 | 27.43 |
| 220 | 27.5 | 04:10:00 | 07:50:00 | 3.73 | 9.82 | 26.18 |
| 230 | 28.75 | 04:05:00 | 07:55:00 | 3.57 | 9.39 | 25.04 |
| 240 | 30 | 04:00:00 | 08:00:00 | 3.42 | 9.00 | 24.00 |

1= It is assumed the ΔV errors perpendicular to the sailcraft's required velocity vector are cancelled either side of the zenith point.

2 = If the laser is applied either side of July 1st then times will be approx. 4mins per solar day later before July 1st and approx. 4mins per solar day earlier after July 1st.

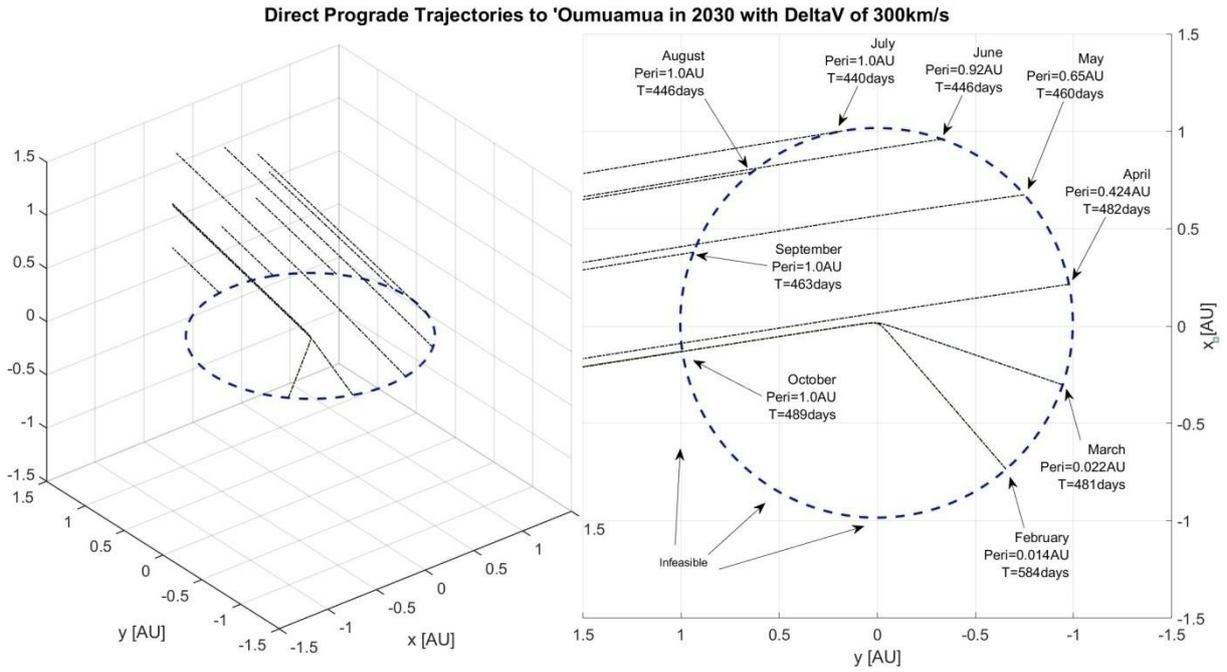

*Figure 1 : Prograde solutions for launching a lasersailspacececraft to 1I/'Oumuamua*

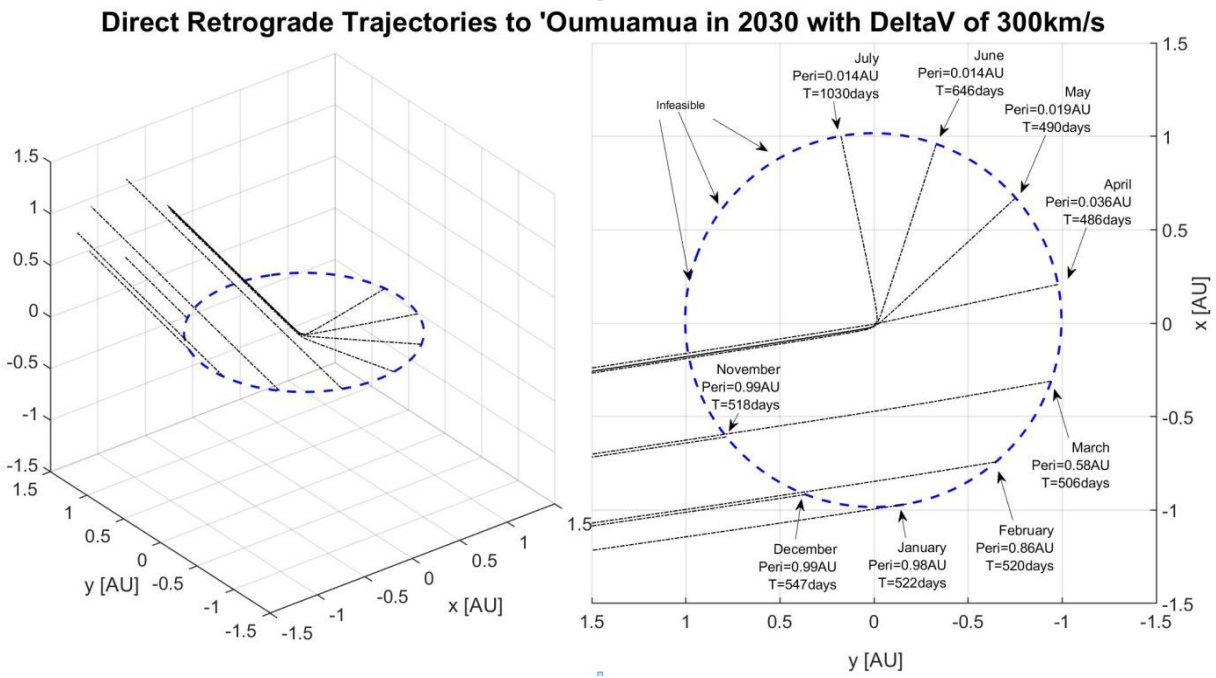

*Figure 2 : Retrograde solutions for launching a lasersailspacececraft to 1I/'Oumuamua*

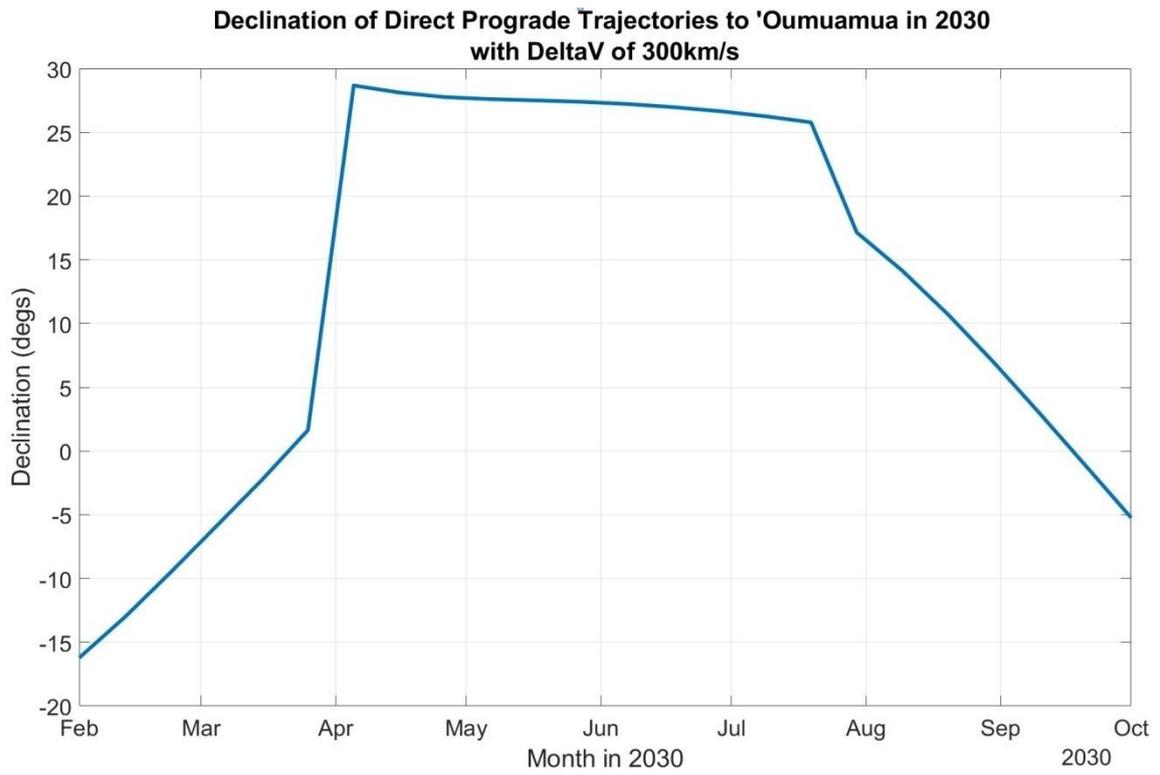

*Figure 3 : Declination for prograde trajectories*

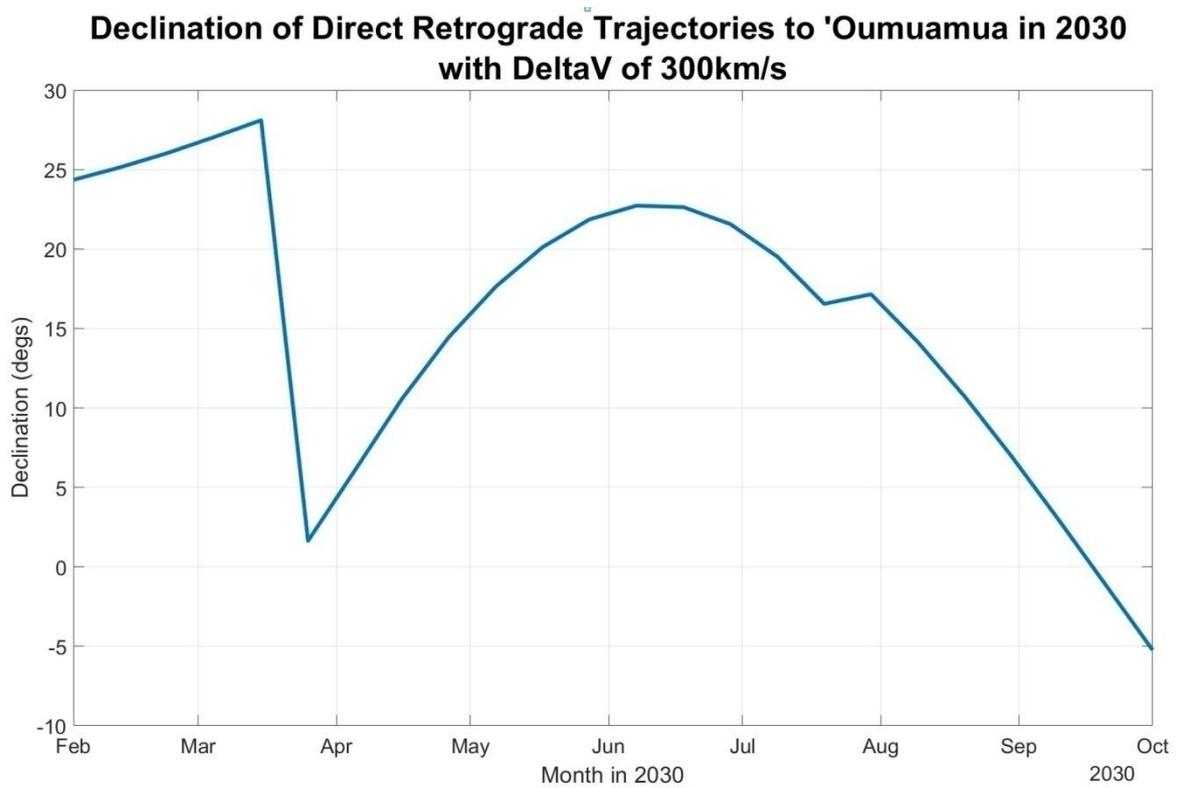

*Figure 4 : Declination for retrograde trajectories*

## 4. Discussion

The results indicate that using an intermediate beaming infrastructure of a much smaller scale (3-30 GW) than for interstellar missions would enable rapid missions to 1I/'Oumuamua. Contrary to the previously proposed mission architectures, based on a combination of flybys, solar Oberth maneuver, and chemical propulsion, using laser sails has several advantages. First, instead of flight times of decades, 1I can be reached within 1-2 years, even for launch dates beyond 2030.

Furthermore, one of the risks of sending a single spacecraft to 1I is that due to the faintness of the object, inaccuracies in the trajectory, and uncertainties in the estimated position of the object, the spacecraft might miss it. The capability of launching multiple spacecraft increases the likelihood that at least one of the spacecraft might obtain useful data. The caveat is obviously that the mass of the spacecraft will be reduced (1-100 kg or even below for lower beaming power). As a consequence, a limited suite of instruments can be carried on an individual spacecraft. In addition, the data rates will also be orders of magnitude lower than for a New Horizons-type spacecraft, which has been proposed before for such missions [37,38]. However, these disadvantages could be balanced by launching individual spacecraft with different instruments and aggregating incoming data. Also, communication technologies, which are currently under development for Breakthrough Starshot might be of use [46], such as using spacecraft swarms as a phased array [47].

According to [40], the cost for the launch (3.3 to 200M$ per 1-100 kg spacecraft, respectively) and infrastructure (450M$ to 4.3B$ for 1-100 kg spacecraft @300 km/s) would be comparable to conventional spacecraft. The infrastructure cost would be amortized over spacecraft launches. Thus, a beamer location enabling launches to a variety of celestial objects would be the most effective way for amortizing infrastructure cost.

## 5. Conclusions

This paper analyzes possible trajectories a sailcraft could take to intercept 'Oumuamua with a launch in 2030. June/July would be optimal in terms of flight time, with a prograde trajectory to maximize the benefit of Earth's orbital velocity (which is around 30km/s), resulting in an overall flight duration of 440days (=1year 2 months).A declination in excess of +20° is indicated for July which would also be approximately the optimum latitude of the laser beam to minimize the thickness of Earth's atmosphere. The acceleration of the sailcraft by the laser beam would be in the morning around 6am.The intercept would take place beyond 82 AU. We conclude that the possibility to launch a large number of spacecraft and reaching 1I much faster than chemical propulsion would remedy several disadvantages of previously proposed mission architectures. It would, therefore, represent an exciting mission for an intermediate beaming infrastructure for Breakthrough Starshot.